\documentclass[prd,twocolumn,showpacs,amsmath,amssymb,floatfix,nofootinbib]{revtex4}
\usepackage{mathrsfs,bm}
\usepackage{longtable,lscape}
\usepackage{txfonts}
\usepackage{amssymb}
\usepackage{indentfirst}
\usepackage{graphicx,,booktabs}
\usepackage{multirow}
\usepackage{color}
\usepackage{amssymb}
\definecolor{cover}{rgb}{0.77,0.87,0.88}
\definecolor{blueone}{rgb}{0.1,0.1,.7}
\definecolor{citec}{rgb}{0.14,0.47,0.09}
\definecolor{two}{rgb}{0.0,0.5,0.}
\definecolor{three}{rgb}{.5,.1,0.15}
\usepackage[bookmarks=true,bookmarksopen=false,plainpages=false,breaklinks=true,
   bookmarksnumbered=true,hypertexnames=false,
   filecolor=blue,urlcolor=three,menucolor=three,
   linkcolor=three,citecolor=blueone, colorlinks,
   anchorcolor=blue,runcolor=pink,frenchlinks=red
   pdfstartview=FitH,pdftitle=title,%
   pdfauthor=author]{hyperref}

\usepackage{relsize}
\usepackage{xspace}
\def\babar{\mbox{\slshape B\kern-0.1em{\smaller A}\kern-0.1em
    B\kern-0.1em{\smaller A\kern-0.2em R}}}

\begin{document}
\title{Hidden-strange molecular states and the $N\phi$ bound state via a QCD  van der Waals force}

\author{Jun He$^{1}$}\email{junhe@njnu.edu.cn}
\author{Hongxia Huang$^{1}$}\email{hxhuang@njnu.edu.cn}
\author{Dian-Yong Chen$^{2}$}
\author{Xinmei Zhu$^{3}$}
\affiliation{$^1$Department of  Physics and Institute of Theoretical Physics, Nanjing Normal University, Nanjing 210097, People's Republic of China\\
$^2$School of Physics, Southeast University, Nanjing 210094, People's Republic of China\\
$^3$Department of Physics, Yangzhou University, Yangzhou, 225009, People's Republic of China}
\date{\today}

\date{\today}
\begin{abstract}
	
In this work, we study the hidden-strange molecular states composed of a baryon and a vector meson in a coupled-channel  $N\rho-N\omega-N\phi-\Lambda K^*-\Sigma K^*$ interaction.  With the help of the effective Lagrangians which coupling constants are determined by the SU(3) symmetry, the  interaction is constructed and inserted into the quasipotential Bethe-Salpeter equation to search for  poles in the complex plane, which correspond to  molecular states. Two poles are found  with a spin parity $3/2^-$ near the $N\rho$ and the $\Sigma K^*$ thresholds, which can be related to the $N(1700)$ and the $N(2100)$, respectively. No pole near the $N\phi$ threshold  can be found if direct interaction between a  nucleon and $\phi$ meson is neglected according to the OZI rule. After introducing  the QCD van der Waals force between a nucleon and $\phi$ meson,  a narrow state can be produced near the $N\phi$ threshold. Inclusion of the QCD van der Waals force changes the line shape of the invariant mass spectrum in
the $N\phi$ channel leading to a worse agreement with the present low-precision data. Future  experiments at BelleII, JLab, and other facilities will be very helpful to clarify the existence of these possible hidden-strange molecular states.

\end{abstract}


\maketitle
\section{INTRODUCTION}

After observations of the $X(3872)$ and the claimed $\Theta$ particle, study of  the exotic hadron becomes a very hot topic in  the hadron physics community~\cite{Choi:2003ue,Nakano:2003qx}.  With  die away of the $\Theta$ particle~\cite{Close:2005pm}, the $XYZ$ particle becomes the most important candidate of  the exotic hadron and great progress is made in both experiment and theory.  Relatively,  the study on the exotic baryon was scarce to some extent until  the observation of two candidates of hidden-charm pentaquarks, $P_c(4450)$ and $P_c(4380)$, at LHCb~\cite{Aaij:2015tga}.  After the LHCb observation,   exotic baryons attract more interest from the hadron physics community. Even some of  five $\Omega_c^*$ baryons  observed recently at LHCb were interpreted as exotic baryons~\cite{Yang:2017rpg,Debastiani:2017ewu,Roca:2015dva,Huang:2018wgr}. 

Now that the candidates of  the pentaquark were observed in the  charmed sector,  possible candidates of the pentaquark in the light sector should be reconsidered. In Refs.~\cite{Yang:2011wz,He:2015cea,He:2016pfa}, the $P_c(4450)$ and $P_c(4380)$ were interpreted as $\Sigma_c \bar{D}^*$ and $\Sigma_c^*\bar{D}$ bound states, respectively. In the light sector,  corresponding hidden-strange pentaquark counterparts were investigated in the molecular state picture in Refs.~\cite{He:2015yva,He:2017aps}.  The results suggest that two states with a spin parity $3/2^-$ can be produced from the $\Sigma K^*$ and $\Sigma^* K$ interactions, which correspond to  the $N(2100)$ observed in the $\phi$ photoproduction and  the $N(1875)$, respectively. Such hidden-strange molecular states are  also supported by analyses of experimental data of relevant  photoproductions as in Refs.~\cite{He:2015yva,He:2017aps,He:2012ud,He:2014gga,Xie:2010yk} and a recent calculation in the constituent quark model~\cite{Huang:2018ehi}.  In an early work in the chiral unitary approach, a pole with the largest coupling to the $\Sigma K^*$ channel was also found at $1977+i53$ MeV from a coupled-channel calculation with a possible spin parity $1/2^-$ or $3/2^-$~\cite{Oset:2009vf}.    

The $N(2100)$ and  LHCb pentaquarks  were observed in  $N\phi$ and $ N J/\psi$ channels, respectively.  In  studies about these pentaquarks in the molecular state picture, especially about their decays, the $N\phi$ and $NJ/\psi$ interactions should be involved. However, we would like to emphasize that in most of the theoretical works the LHCb pentaquarks are not from the  $NJ/\psi $ direct interaction according to the well-known OZI rule, but from a coupled-channel interaction ~\cite{Roca:2016tdh,Xiao:2013yca,Xiao:2016ogq}.  As the $N J/\psi$ interaction,   a quark exchange  between a nucleon and a $\phi$  meson is  also forbidden according to the OZI rule, and often neglected~\cite{Oset:2009vf}. Hence, for studies in the molecular state picture, the interaction between an $s\bar{s}/c\bar{c}$ and a nucleon is often assumed to be very small,  and such kind of the hidden-strange/hidden-charmed  molecular state is often excluded from consideration.

Since a $\phi$ meson can not interact with a nucleon by a quark exchange,  it is a good place to test the effect of  a gluon exchange. In Refs.~\cite{Brodsky:1989jd,Gao:2000az}, the QCD van der Waals force, which reflects a multigluon exchange,  has been suggested to be strong enough to produce a bound state.  Such a proposal is supported by a study in Ref.~\cite{Luke:1992tm}, where it was found that at low velocity the QCD van der Waals interaction is enhanced.   A lattice QCD calculation also supports  the existence of such a kind of a bound state~\cite{Beane:2014sda}.  Hence, the $N\phi$ bound state is a good way to study  the effect of a gluon exchange, that is, the QCD van der Waals force.  Moreover, the $N\phi$ bound state is also a hidden-strange pentaquark. It is interesting  to make a systematic study of possible states from the coupled-channel  $N\rho-N\omega-N\phi-\Lambda K^*-\Sigma K^*$  interaction in both experiment and theory,  which will deepen our understanding about the hidden-strange pentaquark.

Calculations in the constituent  quark model were also performed to study  a possible bound state composed of a nucleon and a $\phi$ meson~\cite{Huang:2005gw,Gao:2017hya}. In all those calculations, the $N\phi$ bound state can be  produced as suggested by Gao $et\ al.$~\cite{Gao:2000az}.   However, in the constituent quark model,  only the one-gluon exchange between two constituent quarks is considered in  the calculation, which contribution vanishes in the $N\phi$ case. The multigluon exchange between two constituent quarks and the gluon exchange with a self interaction are not considered  in these models. Hence,  the attractiveness between a nucleon and $\phi$ meson does not originate from the QCD van der Waals force, which is from a multigluon exchange, but from the delocalization effect in  the quark delocalization color screen model (QDCSM)~\cite{Gao:2017hya} or the $\sigma$ exchange in the chiral quark model~\cite{Huang:2005gw}.   In the constituent quark model, the  $N\phi$ direct interaction is attractive, but not strong enough to produce a bound state and the couplings between $N\phi$ channel. And other channels, such as $\Sigma^* K$ and $\Sigma K^*$~\cite{Gao:2017hya} or $\Lambda K^*$~\cite{Huang:2005gw},  should be introduced to provide enough attractiveness to produce a $N\phi $ bound state. 

In our previous work~\cite{He:2017aps}, we studied the $\Sigma^* K-\Sigma K^*$ interaction, where the coupled-channel effect is very small and the  $\Sigma^* K$ and  $\Sigma K^*$ bound states are almost determined by  corresponding interactions. However, in that work, other channels, such as $N\rho$, $N\omega$ and $\Lambda K^*$, were not included in the coupled-channel calculation. It was done in Ref.~\cite{Oset:2009vf}, and a state with a large coupling to $\Sigma K^*$ was found  at about 1977 MeV, which is consistent  with our conclusion in Ref.~\cite{He:2017aps}.  In the current work, we will make a coupled-channel calculation of $N\rho-N\omega-N\phi-\Lambda K^*-\Sigma K^*$ interaction and extend it to include the $N\phi$ van der Waals force, which will be helpful to understand  the interaction of a nucleon and a $\phi$ meson.

This paper is organized as follows.  In next section,  we present the effective Lagrangians adopted to describe the $N\rho-N\omega-N\phi-\Lambda K^*-\Sigma K^*$ interaction. Corresponding coupling constants
are determined by the SU(3) symmetry.  The QCD van der Waals force is also transformed to a form which can be used in our formalism.  In
Sec.~\ref{Sec: results},  the bound states are searched for and  numerical results are presented. Finally, the paper
ends with  a summary and discussion.

\section{Formalism}\label{Sec: Formalism}

\subsection{$N\rho-N\omega-N\phi-\Lambda K^*-\Sigma K^*$ interaction}\label{Sec: Lagrangian}

In the current work, we consider the channels with a baryon and a vector meson. The potentials for these channels are analogous but with  different coupling constants, which can be related with the help of the SU(3) symmetry. Here, we present first the Lagrangians for the $\Sigma K^*$ interaction as an example. 

In the current work, we will consider both pseudoscalar ($P=\pi$ and $\eta$) and vector ($V=\rho$, $\omega$ and $\phi$) exchanges. To describe the couplings of the $K^*$ meson with exchanged  pseudoscalar and  vector mesons,   we need the Lagrangians as
\begin{align}
	{\cal L}_{K^*K^*V}&=i\frac{g_{K^*K^*V}}{2}( K^{*\mu\dag}{V}_{\mu\nu}K^{*\nu}+K^{*\mu\nu\dag}{V}_{\mu}K^{*\nu}+K^{*\mu\dag}{V}_{\nu}K^{*\nu\mu}),\nonumber\\
	{\cal L}_{K^*K^*P}&=g_{K^*K^*P}\epsilon^{\mu\nu\sigma\tau}\partial^\mu K^{*\nu}
	\partial_\sigma P K^{*\tau}+{\rm H.c.},
\end{align}
where $K^{*\mu\nu}=\partial^\mu K^{*\nu}-\partial^\nu K^{*\mu}$. The flavor structures are  $K^{*\dag}{\bm {A}}\cdot {\bm \tau} K^*$ for an isovector $A$ ($=\pi$ or $\rho$) meson,  and $K^{*\dag}  K^* B$  for an isoscalar $B$ ($=\eta$, $\omega$ or $\phi$) meson. 
The coupling constants can be obtained  from the $\rho\rho\rho$ and $\rho\omega\pi$ couplings with the help of the SU(3) symmetry.  The $g_{\rho\rho\rho}$ is suggested equivalent to $g_{\pi\pi\rho}=6.2$~\cite{Bando:1987br,Janssen:1994uf}.  The SU(3) symmetry suggests $g_{K^*K^*\rho}=g_{K^*K^*\omega}=g_{K^*K^*\phi}/[\sqrt{2}(2\alpha-1)]=g_{\rho\rho\rho}/(2\alpha)$.  For the $K^*K^*P$ vertex, we have $g_{K^*K^*\pi}=g_{K^*K^*\eta}/[-\sqrt{1/3}(1-4\alpha)]=g_{\omega\rho\pi}/(2\alpha)$, and $g_{\omega\pi\rho}=11.2$ GeV$^{-1}$~\cite{Matsuyama:2006rp}.  Here, we adopt $\alpha=1$ for $VVV$ and $VVP$ vertices~\cite{Ronchen:2012eg} .

The Lagrangians for the vertices of  a strange $\Sigma$ baryon and an exchanged vector and peseudoscalar mesons are also required and read
\begin{eqnarray}
	{\cal
	L}_{\Sigma\Sigma V}&
	=&-g_{\Sigma\Sigma V}
	\bar{\Sigma}[\gamma^\nu-\frac{\kappa_{\Sigma\Sigma\rho}}{2m_{\Sigma}}
	\sigma^{\nu\rho}\partial_\rho]{V}^\nu  \Sigma,\nonumber\\
	{\cal
	L}_{\Sigma\Sigma P}&=&-\frac{f_{\Sigma\Sigma P}}{m_\pi}
	\bar{\Sigma}\gamma^5\gamma_\mu \partial^\nu \Sigma.
\end{eqnarray}
The flavor structures are $-i {\bm \Sigma}^\dag \times {\bm \Sigma} \cdot {\bm A}$ and ${\bm \Sigma}^\dag \cdot {\bm \Sigma}B$.   The coupling constants can be obtained with an SU(3) symmetry as $g_{\Sigma\Sigma\rho}=g_{\Sigma\Sigma\omega}=2\alpha g_{NN\rho}$ and $g_{\Sigma\Sigma\phi}=-\sqrt{2}(2\alpha-1) g_{NN\rho}$.  The coupling constant is $g_{NN\rho}=6.1994/2$ as suggested in Ref.~\cite{Matsuyama:2006rp} and 3.02 in Ref.~\cite{Ronchen:2012eg}.  We choose a value of 3.05 here.  In the J\"ulich model,  $\alpha_{BBV}=1.15$~\cite{Ronchen:2012eg}. In the current work, we still adopt the standard value $\alpha_{BBV}=1$. With the SU(3) symmetry, we have relations as  $\kappa_{\Sigma\Sigma\rho}=\kappa_{\Sigma\Sigma\omega}=1/4\kappa_{\rho}$,
$\kappa_{\Sigma\Sigma\phi}=-1/2\kappa_{\rho}$ with $\kappa_\rho=6.1$. For the $VVP$ vertex,  ${f_{\Sigma\Sigma\pi}}=2\alpha{f_{NN\pi}}$ and  ${f_{\Sigma\Sigma\eta}}=\frac{2}{\sqrt{3}}(1-\alpha){f_{NN\pi}}$ with $\alpha_{BBP}=0.4$~\cite{Ronchen:2012eg} and $f_{NN\pi}=1$~\cite{Matsuyama:2006rp, Ronchen:2012eg} .

With the Lagragians above, the potential can be written in a general form,
\begin{align}
	i{\cal V}_{{V}}&=\frac{C^V_{ij}G_V}{q^2-m_{V}^2} 
\bar{B}_i[\rlap\slash A_{ij}-\frac{A_{ij}\cdot q\rlap\slash q}{m_{V}^2}+
\frac{\tilde{\kappa}_{ij}\kappa_\rho(\rlap\slash A_{ij}\rlap\slash q-\rlap\slash q\rlap\slash A_{ij})}{2(m_i+m_j)}
]B_j,\nonumber\\
i{\cal V}_{{P}}
&=\frac{C^P_{ij}G_P}{q^2-m_P^2}i\epsilon^{\mu\nu\sigma\tau}(p'_1+p_1)^\mu \epsilon_i^{\nu\dag}
	q^\sigma \epsilon_j^{\tau}
	~\bar{B}_i\gamma_5\rlap{$\slash$} q  B_j,
\end{align} 
where $A^\mu_{ij}=\epsilon_i^{\dag}\cdot (q-p'_1) \epsilon_j^{\nu}-\epsilon_i^{\dag\nu} \epsilon_j\cdot (q+p_1)+(p_1+p'_1)^\nu~\epsilon_i^{\dag}\cdot\epsilon_j$, $G_V\equiv g_{\rho\rho\rho}g_{NN\rho}/2$ and $G_P=g_{\rho\omega\pi}f_{NN\pi}/m_\pi$. The $i$ and $j$ are for the channel $N\rho$, $N\omega$, $N\phi$, $\Lambda K^*$, or $\Sigma K^*$. Here the $C^{P,V}_{ij}$ coefficient and $\tilde{\kappa}_{ij}$ are listed in  
Table \ref{tab: coefficient}, which are obtained with the SU(3) symmetry as in the $\Sigma K^*$ case.

\renewcommand\tabcolsep{0.39cm} \renewcommand{\arraystretch}{1.5}
\newcommand{\tabincell}[2]{\begin{tabular}{@{}#1@{}}#2\end{tabular}}
\begin{table*}[htbp!]

\caption{The coefficient $C_{ij}$ and $\tilde{\kappa}_{ij}$.  The $\{C^V_{ij},\tilde{\kappa}_{ij}\}_V$ and $\{C^P_{ij}\}_P$  are for vector and pseudoscalar exchanges, respectively.}
\label{tab: coefficient}%
\begin{center}
\begin{tabular}{c|rrrrrrrrrr}
 \toprule[1.5pt]
 & \multicolumn{2}{c}{$N\rho$ }&  \multicolumn{2}{c}{$N\omega$} &  \multicolumn{2}{c}{$N\phi$} &  \multicolumn{2}{c}{$K^*\Lambda$} &  \multicolumn{2}{c}{$K^*\Sigma$ }\\ \hline
$N\rho$ & \multicolumn{2}{c}{$\{{ -2}, 1\}_\rho$ } 
&  \multicolumn{2}{c}{$\{-\sqrt{3}\}_\pi$} & \multicolumn{2}{c}{0}
& \tabincell{r}{$ \{{3\over2}, -{1\over2}\}_{K^*}$ \\  } &  \tabincell{r}{ $\{{1\over2}(1+2\alpha)\}_K$ \\ } 
& \tabincell{r}{$ \{{1\over2}, -{1\over2}\}_{K^*}$ \\  } &  \tabincell{r}{ $\{{1\over2}(2\alpha-1)\}_K$\\ } \\
 $N\omega$  & \multicolumn{2}{c}{\ }   & \multicolumn{2}{c}{0} & \multicolumn{2}{c}{0} 
 & \tabincell{r}{$ \{{-{\sqrt{3}\over2}}, -{1\over2}\}_{K^*}$ \\  } &  \tabincell{r}{ $\{{1\over2\sqrt{3}}(1+2\alpha)\}_K$\\ } 
  & \tabincell{r}{$ \{{{\sqrt{3}\over2}}, -{1\over2}\}_{K^*}$ \\} &  \tabincell{r}{  $\{{\sqrt{3}\over2}(2\alpha-1)\}_K$\\ } \\
 $N\phi$ & \multicolumn{2}{c}{\ } & \multicolumn{2}{c}{\ } & \multicolumn{2}{c}{0} 
   & \tabincell{r}{$ \{{-\sqrt{3\over2}}, -{1\over2}\}_{K^*}$ \\  } &  \tabincell{r}{ $\{-{1\over\sqrt{6}}(1+2\alpha)\}_K$ \\ } 
      & \tabincell{r}{$ \{{\sqrt{3\over2}}, -{1\over2}\}_{K^*}$ \\  } &  \tabincell{r}{ $\{\sqrt{3\over2}(2\alpha-1)\}_K$ \\ } \\
 $K^*\Lambda$ & \multicolumn{2}{c}{\ } & \multicolumn{2}{c}{\ }& \multicolumn{2}{c}{\ }
 & \tabincell{r}{ $\{1, -{1\over4}\}_\omega$ \\ $\{-1,{1\over2}\}_\phi$ } &  \tabincell{r}{   $\{-1+\alpha\}_\eta$\\\ }  
       & \tabincell{r}{$ \{0, 0\}_{\rho}$ \\  } &  \tabincell{r}{$\{(-1+\alpha)\}_\pi$ \\ } \\
 $K^*\Sigma$& \multicolumn{2}{c}{\ } & \multicolumn{2}{c}{\ }& \multicolumn{2}{c}{\ } & \multicolumn{2}{c}{\ } 
 & \tabincell{r}{$\{{-2}, {1\over4}\}_\rho$ \\   $\{1, {1\over4}\}_\omega$ \\ $\{-1,-{1\over2}\}_\phi$ } &  \tabincell{r}{  $\{-2\alpha\}_\pi$ \\  $\{1-\alpha\}_\eta$\\\ }  
 \\
 \bottomrule[1.5pt]
\end{tabular}%
\end{center}
\end{table*}

We would like to show that the potentials obtained by the SU(3) symmetry are comparable to the ones obtained from the chiral Lagrangian in Refs.~\cite{Oset:2009vf, Kolomeitsev:2003kt} after a nonrelativization. Such a comparison was also made in our previous work about $Z_c(3900)$~\cite{He:2015mja} and it was found that the results in our work are consistent with those in the chiral unitary approach~\cite{Aceti:2014uea}. After the nonrelativization, the potential kernel in the current work can be rewritten as
\begin{align}
	i{\cal
	V}_{K^*}
&=\frac{C^V_{ij}G_V}{m_{V}^2} (p^0_i+p^0_j){\bm \epsilon}^\dag_i\cdot {\bm \epsilon}_j.
	\end{align}
In the chiral unitary approach, the corresponding potential is 
\begin{align}
	i{\cal
	V}_{K^*}
&=-\frac{C_{ij}}{4f^2} (p^0_i+p^0_j){\bm \epsilon}^\dag_i\cdot {\bm \epsilon}_j.
	\end{align}
Here $p^0_{(i,j)}$ and ${\bm \epsilon}_{(i,j)}$ are the zero component of the momentum and three-dimensional polarized vector for initial and final mesons. 
The coefficient $C_{ij}$ is the same as those in  Ref.~\cite{Oset:2009vf} if different conventions adopted in the two models are considered.  The coupling constant in our model corresponds to 1.3$f$ in Ref.~\cite{Oset:2009vf}. In the current work, the terms for an anomalous magnetic moment are included in the $BBV$ Lagrangian. Though it will  vanish under a nonrelativization, it is reasonable to assume that the relevant coupling constant will be affected. A similar situation can be found for the  $VVV$ vertex. Besides, as suggested in Ref.~\cite{Oset:1997it}, the $f$ can deviate from the standard value of 93 MeV, and the increase of the $f$ can be compensated by the increase of the cutoff. Hence, in the current work, we still choose the coupling constant obtained with the $g_{\rho\rho\rho}$, $g_{NN\rho}$, $g_{\rho\omega\pi}$, and $f_{NN\pi}$.  

\subsection{QCD van der Waals force}\label{Sec: Waals}

As suggested by Gao $et\ el.$~\cite{Gao:2000az}, a $\phi$ meson interacts with a nucleon with the QCD van der Waals force, which was adopted as a nonrelativistic Yukawa-type attractive potential of a form $V_{s\bar{s},N}=-\alpha e^{-\mu r}/r$.  The parameters were chosen as $\alpha=1.25$ and $\mu=0.6$ GeV, which are consistent with the values  for the $c\bar{c}$ charmonium adopted by Brodsky $et\ al.$~\cite{Brodsky:1989jd}. We would like to remind that as suggested in  Ref.~\cite{Brodsky:1989jd} there should be other spin-orbit and spin-spin hyperfine terms because the interaction is vectorlike but the parameters are determined with this simple form of the potential. Here we only keep the main part as in  Refs.~\cite{Brodsky:1989jd,Gao:2017hya}, which ensures that the discussion about the parameters in those references can be adopted here.   In the current work, we work in the momentum space,  the QCD van der Waals force above should be transformed to an Yukawa-type interaction as,
\begin{align}
i{\cal V}_{N\phi}=(4\pi)\frac{\alpha}{q^2-\mu^2} 2m_N \bar{N} N~\phi\cdot\phi,
\end{align}
where $N$ and $\phi$ are the spinor for a nucleon and polarized vector for a $\phi$ meson, respectively. An additional $2m_N$ is introduced due to the convention adopted in our formalism. 

\subsection{Quasipotential Bethe-Salpeter equation}\label{Sec: BS}

With the potentials presented above, the $N\rho-N\omega-N\phi-\Lambda K^*-\Sigma K^*$ interaction can be inserted into the Bethe-Salpeter equation to find  molecular states. Because of  the difficulty in solving the 
Bethe-Salpeter equation in Minkovsik space, as in previous work~\cite{He:2012zd,He:2013oma,He:2014nya,He:2015cca}, a spectator quasipotential approximation will be introduced by putting one of the two particles
on shell~\cite{Gross:2008ps,Gross:2010qm}. In Ref.~\cite{Gross:1999pd},  the author suggested that the heavier particle should be put on shell when a one-boson
exchange is adopted while in Ref~\cite{He:2017aps}, a test of different choices of 
on shell particle was made and its effect on numerical  results were found small.  The method was explained explicitly in the Appendixes of Ref.~\cite{He:2015mja}. In this work, we will put the heavier particle on shell.

A bound state produced from the $N\rho-N\omega-N\phi-\Lambda K^*-\Sigma K^*$ 
interaction is reflected by a pole of  scattering amplitude ${\cal
M}$. The  quasipotential Bethe-Salpeter equation for a partial-wave
amplitude with a fixed spin parity $J^P$ reads
~\cite{He:2015mja,He:2015cea}
\begin{eqnarray}
i{\cal M}^{J^P}_{\lambda'\lambda}({\rm p}',{\rm p})
&=&i{\cal V}^{J^P}_{\lambda',\lambda}({\rm p}',{\rm
p})+\sum_{\lambda''\ge0}\int\frac{{\rm
p}''^2d{\rm p}''}{(2\pi)^3}\nonumber\\
&\cdot&
i{\cal V}^{J^P}_{\lambda'\lambda''}({\rm p}',{\rm p}'')
G_0({\rm p}'')i{\cal M}^{J^P}_{\lambda''\lambda}({\rm p}'',{\rm
p}).\quad\quad \label{Eq: BS_PWA}
\end{eqnarray}
The partial-wave potential with a  fixed spin parity $J^P$ is obtained from the potential kernel ${\cal V}_{\lambda'\lambda}$ in  previous section as
\begin{eqnarray}
i{\cal V}_{\lambda'\lambda}^{J^P}({\rm p}',{\rm p})
&=&2\pi\int d\cos\theta
~[d^{J}_{\lambda\lambda'}(\theta)
i{\cal V}_{\lambda'\lambda}({\bm p}',{\bm p})\nonumber\\
&+&\eta d^{J}_{-\lambda\lambda'}(\theta)
i{\cal V}_{\lambda'-\lambda}({\bm p}',{\bm p})],
\end{eqnarray}
where  initial and final relative
momenta are chosen as ${\bm p}=(0,0,{\rm p})$  and ${\bm p}'=({\rm
p}'\sin\theta,0,{\rm p}'\cos\theta)$ with a definition ${\rm
p}^{(')}=|{\bm p}^{(')}|$, and $d^J_{\lambda\lambda'}(\theta)$ is the
Wigner d matrix.  These equations can be extended to a coupled-channel case as in Ref.~\cite{He:2014nxa}.

A regularization should be introduced in the  integral equation  to make it convergent. In our previous works~\cite{He:2014nya,He:2015cca}, an  exponential regularization is introduced  in the propagator of a form 
$ G_0({\rm p})\to G_0({\rm
 p})\left[e^{-(p_1^2-m_1^2)^2/\Lambda^4}\right]^2$,  
where $k_1$ and $m_1$ are  momentum and the mass of the off shell meson,
respectively.
The exponential
regularization is a softer version of the cutoff regularization  in the chiral unitary approach, and the cutoff $\Lambda$ used here plays a similar role to the cutoff ${\rm p}_{max}$  in the chiral unitary approach~\cite{Oset:2009vf}. The interested reader is referred to Ref.~\cite{He:2015mja} for
further information about the regularization.  In this work, we will consider both exponential and cutoff regularizations.

Now that the regularization  ensures the convergence of  an integral equation, combined with a discussion in Ref.~ \cite{Lu:2016nlp},  a form factor is not essential to be introduced for the exchanged meson, which was applied in that work and our previous work in Ref.~\cite{He:2016pfa}. In the current work, we follow such a treatment.

The integral equation can be transformed to a matrix equation by discreting the momenta ${\rm p}$,
${\rm p}'$, and ${\rm p}''$ by the Gauss quadrature with wight $w({\rm
p}_i)$ as
\begin{eqnarray}
{M}_{ik}
&=&{V}_{ik}+\sum_{j=0}^N{ V}_{ij}G_j{M}_{jk},
\end{eqnarray}
where the indices for channel, helicity and  momentum are all absorbed into an index $i$. The  propagator is now of a form,
\begin{eqnarray}
	G_{j>0}&=&\frac{w({\rm p}''_j){\rm p}''^2_j}{(2\pi)^3}G_0({\rm
	p}''_j), \nonumber\\
G_{j=0}&=&-\frac{i{\rm p}''_o}{32\pi^2 W}+\sum_j
\left[\frac{w({\rm p}_j)}{(2\pi)^3}\frac{ {\rm p}''^2_o}
{2W{({\rm p}''^2_j-{\rm p}''^2_o)}}\right],
\end{eqnarray}
where the ${\rm p}_o=\lambda(W, M_j, m_j)$  is the on shell  momentum with $\lambda(x,y,z)=\sqrt{(x^2-(y+z)^2)(x^2-(y-z)^2)}/2x$ and $W$, $M_j$, $m_j$ being the meson-baryon energy, and the masses of two particles in $j$ channel. 
Here, we would like to note that  there only exists the on shell term (with $j=0$) at energies under the corresponding threshold. 

The pole can be searched by the variation of $z$ to satisfy $|1-V(z)G(z)|=0$, 
where  $z=E_R+i\Gamma/2$ equals the meson-baryon energy $W$ at the real axis. 
  The scattering amplitude where the initial and final particles are on shell is
\begin{eqnarray}
	M=M_{00}=\sum_j[(1-{ V} G)^{-1}]_{0j}V_{j0}.\label{Eq: Scattering ampltiudes}
\end{eqnarray}
The above equation indicts that the scattering amplitude is meaningless at energies under the threshold in our approach.

\section{Numerical results}\label{Sec: results}
With the above preparation, the molecular states from the $N\rho-N\omega-N\phi-\Lambda K^*-\Sigma K^*$  interaction can be searched in the complex plane of $z$. For the $N\phi$ interaction with  the QCD van der Waals force, no pole can be found in the case of a spin parity $1/2^-$.  Hence, in this work, we focus on the case of a spin parity  $3/2^-$, which is in an S-wave and was studied in the constituent quark model~\cite{Gao:2017hya}. 

\subsection{States from $N\rho-N\omega-N\phi-\Lambda K^*-\Sigma K^*$  interaction}\label{Sec: wo}

\begin{figure}[h!]
\begin{center}
\includegraphics[bb=88 150 350 300, clip, scale=0.94]{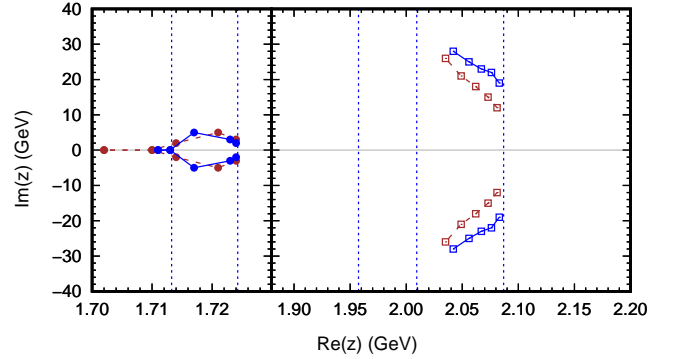}
\end{center}
\caption{The poles  from the $N\rho-N\omega-N\phi-\Lambda K^*-\Sigma K^*$  interaction.  The points linked by a full line (brown) and those by a dashed line (blue) are for the pole with an exponential regularization at a cutoff $\Lambda$ from 0.9 to 1.3 GeV with a step of 0.1 GeV, and for the pole with cutoff regularization at a cutoff ${\rm p}_{max}$ from 0.55 to 0.75 GeV with a step of 0.05 GeV, respectively. The explicit explanation can be found in the text. The vertical dashed (blue) lines are for thresholds of the  $N\rho$, $N\omega$, $N\phi$, $\Lambda K^*$, $\Sigma K^*$ channels from left to right.
}
\label{Fig: Lambda}
\end{figure}

First we consider the interaction without the QCD van der Waals force and the results are given in Fig.~\ref{Fig: Lambda}. Here, the poles from the interaction with both the exponential and cutoff regularizations are illustrated in Fig.~\ref{Fig: Lambda}. We choose five values of the cutoff  for tthe wo regularizations, respectively, and a pole at five different cutoffs is presented  as five points linked by a line.   

In the energy region from 1.7 to 2.20 GeV, which covers the thresholds of  the five channels considered in the current work, two poles can be generated from the interaction.  The higher one is at an energy about 2050 MeV near the $\Sigma K^*$ threshold. Five points for this pole are obtained at $\Lambda=0.9$, $1.0$, $1.1$, $1.2$, and $1.3$ GeV for the exponential regularization, and at ${\rm p}_{max}=0.55$, $0.60$, $0.65$, $0.70$ and $0.75$ GeV from right to left , respectively.  For both regularizations, with the increase of the cutoff, the real and imaginary parts of the
 position of this pole decrease and increase, respectively. It means that the mass and width of  the corresponding molecular state become larger and smaller, respectively, with the increasing of the cutoff. For the points at the same energy, the one  with the exponential regularization has a larger width than that with a cutoff regularization.  

Another pole is predicted at about 1710 MeV. Because the $N\rho$ channel is the lowest channel considered in our work, the pole is a bound state at a real axis at $\Lambda=1.2$ or $1.3$ GeV with an exponential regularization  and at a ${\rm p}_{max}=0.70$ or $0.75$ GeV with a cutoff regularization. With the decrease of the cutoff, the pole runs  to the $N\rho$ threshold. After crossing the threshold, the pole deviates from the real axis and the width increases first then decreases with the increase of  the cutoff. Different from the higher pole, the lower pole almost runs in the same path in the two cases with exponential and cutoff regularizations.  Generally speaking, both regularizations give consistent results if an appropriate cutoff is chosen. In the following, we only give the results with the cutoff regularization.

In Sec.~\ref{Sec: Lagrangian}, after nonrelativization, the main difference between the potential in the current work and  that in the chiral unitary approach adopted in Ref.~\cite{Oset:2009vf} is  different values of the coupling constants. If we choose  a coupling constant which corresponds to a standard value of $f=93$ MeV as in Ref.~\cite{Oset:2009vf}, the higher pole  is found at  $2024+32i$ MeV at a ${\rm p}_{max}=0.55$ GeV or at  $1973+24i$ MeV at a ${\rm p}_{max}=0.75$ GeV.  It is consistent with the conclusion that the increase of the value of $f$ can be compensated by increasing  the cutoff as suggested in  Ref.~\cite{Oset:1997it}.

\begin{figure}[h!]
\begin{center}
\includegraphics[bb=80 50 370 300, clip, scale=0.87]{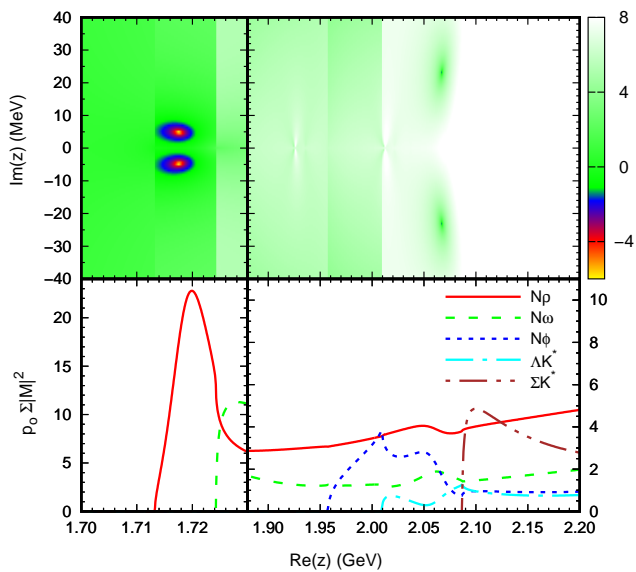}
\end{center}
\caption{$\log|1-V(z)G(z)|$ with the variation of  $z$ for the $N\rho-N\omega-N\phi-\Lambda K^*-\Sigma K^*$   interaction at a ${\rm p}_{max}=0.65$ GeV without the QCD van der Waals force (upper panel), and corresponding ${\rm p}_o\sum_{\lambda\lambda'} |M_{\lambda\lambda'}|^2$ for $N\rho$ (red full line), $N\omega$ (green dashed line), $N\phi$ (blue dotted line),  $\Lambda K^*$ (cyan dash-dotted line), and $\Sigma K^*$ (brown dash-dot-dotted line) channels with the variation of Re$(z)$ (lower panel).
}
\label{Fig: plot}
\end{figure}

In Fig.~\ref{Fig: plot}, the explicit results about the two poles are presented.  Here we take  result at cutoff ${\rm p}_{max}=0.65$ GeV as an example. 
 Two poles can be observed obviously from the $\log|1-V(z)G(z)|$ in the complex plane of $z$.  In the experiment, a state is usually observed in the invariant mass spectrum. Here we take the square of the amplitudes  ${\rm p}_o\sum_{\lambda\lambda'} |{\cal M}_{\lambda\lambda'}|^2$  as the invariant mass spectrum with ${\cal M}$ and  $p_o$ being the on shell amplitude and momentum of  the corresponding channel. The invariant mass spectra  for  the $N\rho$, $N\omega$, $N\phi$,  $\Lambda K^*$, $\Sigma K^*$ channels  are also presented in Fig.~\ref{Fig: plot}.  Under the threshold of a channel, the  amplitude is meaningless and can not be calculated in our model as shown in Eq.~(\ref{Eq: Scattering ampltiudes}). It makes  it difficult to identify  the strength of  the couplings of a state to all channels  as in Ref.~\cite{Oset:2009vf}. However, the invariant mass spectrum still provides some interesting information. Corresponding to the lower pole, an obvious resonance structure can be observed in the $N\rho$ channel, which reflects that the coupling of this state to the $N\rho$ channel is considerable.  An obvious enhancement is also found near the $N\omega$ threshold.  For the higher pole,  no obvious standard resonance structure can be observed from our results. However,  the existence of a higher pole affects the shapes of the invariant mass spectra for the $N\phi$, $\Lambda K^*$ channels obviously.  The effects on the $N\rho$ and $N\omega$ channels are also observable but relatively smaller.  Though the invariant mass spectrum for the $\Sigma K^*$ channel at the mass of the pole can not be provided, a sharp increase near the $\Sigma K^*$ threshold suggests  that the effect of this pole is large on the $\Sigma K^*$ channel.  Besides, a cusp can be found at the $\Lambda K^*$ threshold in the $N\phi$ channel.

\subsection{States with QCD van der Waals force}\label{Sec: w}

\begin{figure}[h!]
\begin{center}
\includegraphics[bb=88 150 350 300, clip, scale=0.94]{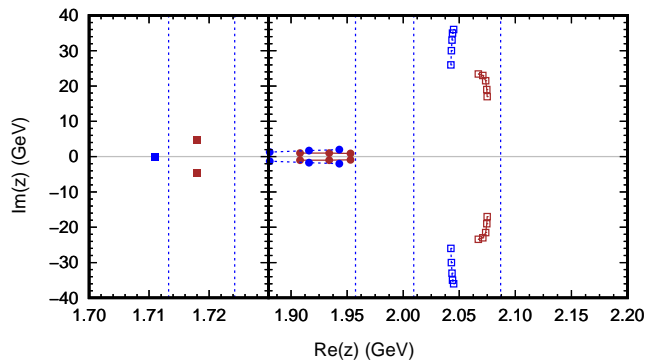}
\end{center}
\caption{The poles from  the $N\rho-N\omega-N\phi-\Lambda K^*-\Sigma K^*$  interaction with the QCD van der Waals force.  The poles linked by a full line (brown) and those by a dashed line (blue) are the results at a cutoff ${\rm p}_{max}=0.65$ GeV, and results at a cutoff ${\rm p}_{max}=0.75$ GeV, respectively. The vertical lines are thresholds as in Fig~\ref{Fig: Lambda}.
}
\label{Fig: Lambdaphi}
\end{figure}

Now we turn to the case with the QCD van der Waals force  included in the  $N\phi$ channel.  The QCD van der Waals force is not well determined with the existent experimental and theoretical information. In Ref.~\cite{Gao:2000az}, authors suggested a strength of $\alpha=1.25$. In this work, we will vary the parameter $\alpha $  to present the results of the QCD van der Waals force with different strength parameters $\alpha$=0, 0.4, 0.8, 1.2, and 1.6, which  are illustrated in Fig.~\ref{Fig: Lambdaphi}.

To show the effect of different cutoffs,  we provide the results at two cutoffs, ${\rm p}_{max}=0.65$ and $0.75$ GeV.  At first sight, one can find that the inclusion of the QCD van der Waals force in the $N\phi$ channel affects the two poles shown in Fig.~\ref{Fig: Lambda} slightly at both cutoffs. The position of the lower pole is almost unchanged, and the higher pole moves from the original position a little. 
Though the effect on two poles produced without the QCD van der Waals force are small, an additional pole close to the real axis will be generated near $N\phi$ threshold with a strength $\alpha=1.2$. With the  increasing of the strength, the pole  runs to a lower energy.  The imaginary part of the position of  this pole is about 1 MeV, which means that the corresponding resonance structure is very narrow.

\begin{figure}[h!]
\begin{center}
\includegraphics[bb=80 50 370 300, clip, scale=0.87]{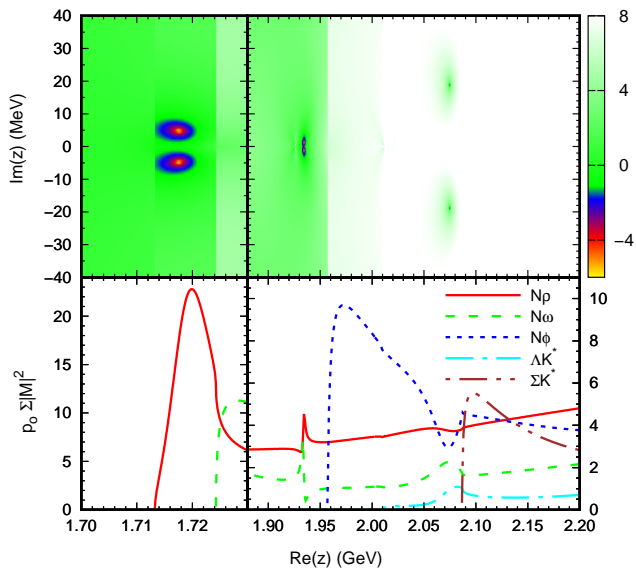}
\end{center}
\caption{$\log|1-V(z)G(z)|$ with the variation of  $z$ for the $N\rho-N\omega-N\phi-\Lambda K^*-\Sigma K^*$ interaction at ${\rm p}_{max}=0.65$ GeV with the QCD van der Waals force  (upper panel), and corresponding ${\rm p}_o\sum_{\lambda\lambda'} |M_{\lambda\lambda'}|^2$. The notation is as Fig.~\ref{Fig: plot}.
}
\label{Fig: plotphi}
\end{figure}

The explicit results at cutoff ${\rm p}_{max}=0.65$ GeV and $\alpha$=1.2  are presented in Fig.~\ref{Fig: plotphi} as an example. Three poles can be observed obviously from the $\log|1-V(z)G(z)|$ in the complex plane of $z$.  Compared with Fig.~\ref{Fig: plot}, the pole and the invariant mass spectra are almost the same at the lower energy region.  At the higher energy region, a sharp peak appears in the invariant mass spectra in the $N\rho$ and $N\omega$ channels. Except for this peak, the line shape for these two channels are almost unchanged as well as that for the $\Sigma K^*$ channel. Obvious change can be found in the $N\phi$ channel, where the QCD van der Waals force is included. An obvious enhancement appears near the $N\phi$ threshold, which is obviously relevant to the existence of the new pole at $1934+1i$ MeV.     

\subsection{Comparison with experiment}\label{Sec: belle}

\begin{figure}[h!]
\begin{center}
\includegraphics[bb=65 55 390 300, clip, scale=0.74]{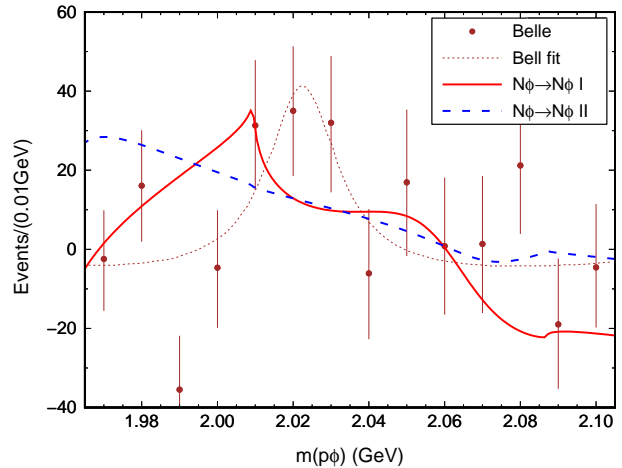}
\end{center}
\caption{The invariant mass spectrum in $N\phi$ channel.  The theoretical results of $k k' \sum_{\lambda\lambda'} |M_{\lambda\lambda'}|^2$ are compared with the Belle experiment~\cite{Pal:2017ypp}. The full (red) and dashed (blue) lines are for the results without and with the QCD van der Waals force. The dotted (brown) line is for the best fitted result by the Belle Collaboration. 
}
\label{Fig: Belle}
\end{figure}

Up to now, there is no direct experimental evidence  for the hidden-strange pentaquark. Recently, a search for the possible hidden strange partner of the LHCb pentaquark was done at Belle in the $N\phi$ invariant mass spectrum~\cite{Pal:2017ypp}. The Belle Collaboration concluded that no evidence was seen for a hidden-strangeness pentaquark. Here, we would like to make a discussion of our results compared with the Belle experiment. Since  the experimental data are not so precise, we only make a qualitative discussion.  The comparison between the invariant mass spectrum $p_o p'_o \sum_{\lambda\lambda'} |M_{\lambda\lambda'}|^2$ with $p'_o=\lambda(\Lambda_c,m(N\phi), m_\pi)$ and the Belle data are presented in Fig.~\ref{Fig: Belle}.

Though no evidence for the hidden-strange pentaquark is observed,  Belle still gives the best fitted result which we present in Fig.~\ref{Fig: Belle} also. It suggests a resonance at 2025 MeV with a width of 22 MeV~\cite{Pal:2017ypp}.  If the QCD van der Waals force is not included, a cusp is produced at the $\Lambda K^*$ threshold, and the pole at 2075+$i$19 MeV exhibits as a shoulder at energies about 2050 MeV.  After including the  QCD van der Waals force in the $N\phi$ channel, there is large enhancement near the $N\phi$ threshold (see Fig.~\ref{Fig: plotphi}).  It seems that the $N\phi$ invariant mass spectrum without the QCD van der Waals force is more consistent with the best fitted line given by the Belle collaboration~\cite{Pal:2017ypp}. In Ref.~\cite{Xie:2017mbe}, the authors also found that the peak in experiment could be from a 	
triangular singularity, but not from the $P_s$ state. Considering the large uncertainty of the data, no conclusion can be obtained from the current experimental data and the theoretical results. However, there exists obvious difference between the results with and without the QCD van der Waals force included, which can be checked by high-precision data.

\section{Summary and discussion}\label{Sec: summary}

In this work, we study the hidden-strange pentaquarks from a coupled $N\rho-N\omega-N\phi-\Lambda K^*-\Sigma K^*$ interaction. The interaction is described with the help of the effective Lagrangians and  coupling constants determined with the SU(3) symmetry. The obtained potential is consistent with those in the chiral unitary approach in Ref.
~\cite{Oset:2009vf} after nonrelativization but with smaller couplings. The potential is inserted into the quasipotential Bethe-Salpeter equation to find thte poles corresponding to the hidden-strange molecular states.

Two poles can be observed in the complex plane for the $N\rho-N\omega-N\phi-\Lambda K^*-\Sigma K^*$ interaction. The higher one is under the $\Sigma K^*$ threshold, which has a large effect on the $\Sigma K^*$ invariant mass spectrum.  This pole also exhibits itself as a shoulder in the $N\phi$ channel, where a cusp appears at the $\Lambda K^*$ threshold. A lower pole can be found near the $N\rho$ threshold. The corresponding resonance structure can be seen obviously in the $N\rho$ channel.  These two poles can be related to the poles at $1977+i53$ MeV and $1695$ MeV suggested in the chiral unitary approach in Ref.~\cite{Oset:2009vf}. Our result is consistent with their conclusion that these two states have the largest coupling to $\Sigma K^*$ and $N\rho$, respectively. In Ref.~\cite{Oset:2009vf}, the spin parity is not well determined while in the current work both states are from the interaction with a spin parity $3/2^-$. 

We also extend our results to include the  QCD van der Waals force  in the $N\phi$ channel. The original two poles are affected a little by the  QCD van der Waals force.  An additional pole near the $N\phi$ threshold is generated after the  QCD van der Waals force is included. Both results of the pole and the invariant  mass spectrum suggest that the couplings of this state to other channels are small, which makes the peak in the invariant mass spectrum  very narrow.  The $N\phi$ bound state was also found very narrow in the constituent quark model~\cite{Gao:2017hya}. It is noteworthy that there may be also the QCD van der Waals force  in other channels if it exists in the $N\phi$ channel. It needs more theoretical efforts to clarify the origin and effect of such a QCD van der Waals force.

Up to now, there is no direct evidence for a hidden-strange pentaquark in experiment. As suggested in  Refs.~\cite{Oset:2009vf,He:2015yva,He:2017aps}, two poles produced from the $N\rho-N\omega-N\phi-\Lambda K^*-\Sigma K^*$ interaction without the QCD van der Waals force can be assigned as  the $N(1700)3/2^-$ and $N(2100)3/2^-$. The results in the current work and  in the constituent quark model~\cite{Gao:2017hya} suggests that the $N\phi$ state is a narrow state, which should be easy to observe in experiment. Besides, the
$N\phi$ invariant mass spectra change obviously. So far, the inclusion
of the QCD Van der Waals force leads to a worse agreement with the present
low-precision data.  Hence, the search of such a state in an experiment at the facilities, such as BelleII, JLab, with improved
statistics is strongly suggested.  The experimental research and further theoretical study will be helpful to clarify this interesting issue.

\section*{Acknowledgments}

This project is partially supported by the National
Natural Science Foundation of China (Grants No.11675228, No. 11375240, 11675080, and No. 11775050) , the Major State Basic Research
Development Program in China (Grant No. 2014CB845405), and the Natural Science
Foundation of the Jiangsu Higher Education Institutions of
China (Grant No. 16KJB140006)

\end{document}